\documentclass[amsmath, amsfonts, showpacs, twocolumn, prl]{revtex4}
\usepackage{graphicx}
\usepackage{epsfig}
\usepackage{bm}
\usepackage{euscript}
\usepackage{dcolumn}
\usepackage{amsmath}
\usepackage{amssymb}
\usepackage[varg]{txfonts}

\def\t{\mathbf{t}}
\def\r{\mathbf{r}}
\def\Tr{\mathrm{Tr}}

\begin{document}

\title{Coulomb drag in quantum circuits}
\author{Alex Levchenko}
\author{Alex Kamenev}

\affiliation{Department of Physics, University of Minnesota,
Minneapolis, MN 55455, USA}

\begin{abstract}
We study drag effect in a system of two electrically isolated
quantum point contacts (QPC), coupled by Coulomb interactions. Drag
current exhibits  maxima as a function of QPC gate voltages when the
latter are tuned to the transitions between quantized conductance
plateaus. In the linear regime this behavior is due to enhanced
electron-hole asymmetry near an opening of a new conductance
channel. In the non-linear regime the drag current is proportional
to the shot noise of the driving circuit, suggesting  that the
Coulomb drag experiments may be a convenient way to measure the
quantum shot noise. Remarkably, the transition to the non-linear
regime may occur at driving voltages substantially smaller than the
temperature.
\end{abstract}

\date{October 22, 2008}


\maketitle

\begin{figure}
  \includegraphics[width=8cm,height=9cm]{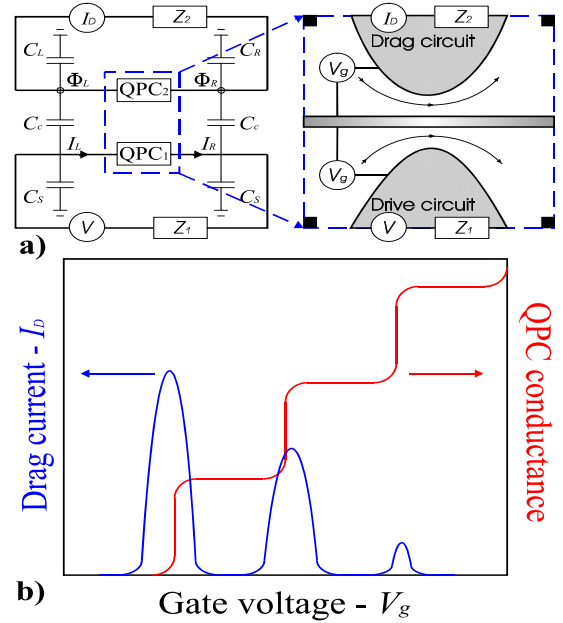}
  \caption{(Color online)
\textbf{a)} Two coupled QPCs and surrounding electric circuitry. The
Coulomb coupling is due to mutual capacitances $C_c$.  Gate voltage
$V_{g}$  control transmission of  e.g.  drive  QPC.   \textbf{b)}
Schematic representation of linear conductance of the drive QPC
along with the drag current as a function of the gate voltage.
}\label{Fig1}
\end{figure}

Drag effect in bulk 2D systems is well established
experimentally~\cite{Solomon,Gramila,Sivan,Lilly,Pillarisetty,Savchenko}
and studied
theoretically~\cite{Smith,MacDonald,Kamenev-Oreg,Flensberg}. By now
it is one of the standard ways to access and measure
electron--electron scattering. Very recently a number of experiments
were performed to study Coulomb drag in quantum confined geometries
such as quantum wires~\cite{Debray-1,Debray-2,Morimoto,Yamamoto},
quantum dots~\cite{Aguado,Kouwenhoven} or quantum point contacts
(QPC)~\cite{Khrapai}. In  these systems a source-drain voltage $V$
is applied to generate current in the \textit{drive  circuit} while
an induced current (or voltage) is measured in the \textit{drag
circuit}. Such a drag current  is a function of the drive voltage
$V$ as well as gate voltages, which controls transmission of one or
both circuits. Figure~\ref{Fig1}a shows  an example of such a setup,
where both drive and drag circuits are represented by two QPC's.

It was reported
~\cite{Debray-1,Debray-2,Morimoto,Kouwenhoven,Khrapai} that the drag
current exhibits maxima for specific values of the gate voltage,
where the drive QPC is tuned to an opening of another conductance
channel. This observation is depicted schematically in
Fig.~\ref{Fig1}b. It was attributed to the shot noise of the drive
QPC~\cite{Kouwenhoven,Khrapai,Aguado,Chudnovskiy}, which is
known~\cite{Lesovik,Reznikov} to exhibit a qualitatively similar
behavior. The idea is that the drag circuit serves as a detector and
a rectifier of the {\em quantum} shot noise in the drive circuit.
Although plausible and in a certain regime indeed correct, this
mechanism differs substantially from the one familiar from the bulk
2D  drag effect. In the latter case drag may be interpreted
\cite{Kamenev-Oreg} as a rectification of nearly equilibrium  {\em
classical} thermal fluctuations in the drive circuit. As a result
the drag current is a power-law function of the temperature ($\sim
T^2$ in many cases~\cite{Footnote-1}). Such a rectification is only
possible due to electron-hole asymmetry in both circuits (otherwise
drag currents of electrons and holes cancel each other). In the bulk
systems the asymmetry is due to a small curvature of the particles
dispersion relation near the Fermi energy.

Mesoscopic  and quantum circuits with the spatial dimensions less
than the temperature length $L_T=v_F/T$ and voltage length
$L_V=v_F/eV$ differ from the bulk 2D systems in several important
ways.  (i) The electron-hole symmetry in such  devices is broken
much stronger than in bulk systems. In mesoscopic devices this is
due to a random configurations of impurities~\cite{Aleiner}, while
in the QPC's the effect is due to the energy dependence of
transmission coefficients. Because of the latter the hole's
transmission probability is typically less than that of the
electrons. (ii) The spatial inversion symmetry may be broken by
left-right asymmetry of the circuit design. As we show below, this
makes two polarities of the drive voltage $V$ to be essentially
non-equivalent. (iii) Because of the above, the quantum circuits may
be easily driven out of the linear response domain (unlike the bulk
systems). Typically a voltage needed to drive a quantum circuit into
a nonlinear regime is parametrically smaller than the temperature.

In this paper we study drag effect between two QPC's,
Fig.~\ref{Fig1}a. We assume  weak interaction between the two
circuits mediated by mutual capacitances $C_c$, Fig.~\ref{Fig1}a.
Since the external circuits typically include also dissipative
elements, the actual interaction is, in general, frequency-dependent
\cite{Aguado} and determined by a {\em matrix} of trans-impedances
(see below).  Because of the weak coupling the drag current $I_D$ is
small and therefore the drag circuit is assumed to be close to
equilibrium. On the other hand, the drive circuit may be
substantially out-of-equilibrium, due to an applied bias $V$. With
these assumption we evaluate the drag current $I_D$ in  the
second-order in the inter-circuit interactions in the framework of
the Keldysh diagrammatic technique (to account for non-equilibrium
conditions of the drive circuit). Details of the calculations are
reported as a supplementary material.

We show that at sufficiently small driving voltage $V$ the drag
current is linear $I_D\propto V$. In this regime the mechanism of
the drag is similar to that in the bulk 2D systems: i.e.
rectification of near-equilibrium thermal noise. Consequently
$I_D\propto T^2$ at small temperatures. The rectification relies on
the electron-hole asymmetry, which is due to energy dependence of
the transmission probability in a given channel. The asymmetry is
the strongest near an opening of a new conductance channel. Indeed,
in this case thermally excited electrons are much more likely to be
transmitted than the holes. Hence the behavior sketched in
Fig.~\ref{Fig1}b (though with no relation to the quantum shot
noise). At larger drive voltages $I_D\propto V^2$ and the effect is
indeed due to the detection of the \textit{excess} shot noise of the
drive circuit~\cite{Kouwenhoven,Khrapai,Aguado,Chudnovskiy}. The
energy dependence of the transmission probability is not required in
this regime, and $I_D$ is proportional to the celebrated Fano
factor~\cite{Lesovik,Reznikov,BB}. Remarkably, the crossover between
the two regimes takes place at $eV\sim T^2/\Delta \ll T$, where
$\Delta$ is an energy scale of the transmission probability.


Quantitatively we found the following expression for the drag
current:
\begin{equation}\label{I-D-trace}
I_D(V)=
\int\!\frac{d\omega}{4\pi\omega^{2}}\mathrm{Tr}\left[\hat{\mathcal{Z}}(\omega)
\hat{S}_{1}(\omega,V)\hat{\mathcal{Z}}(-\omega)
\hat{\Gamma}_{2}(\omega)\right]\,.
\end{equation}
Here and throughout the paper indexes $1,2$ refer to the drive and
drag circuits, correspondingly. The elements
$\mathcal{Z}_{ab}(\omega)$, $a,b=R,L$ of the trans-impedance matrix
$\hat{\mathcal{Z}}(\omega)$ encode inter-circuit coupling. They are
defined as
$\mathcal{Z}_{ab}(\omega)=\partial\Phi_{a}(\omega)/\partial
I_{b}(\omega)$, where the corresponding local fluctuating currents
$I_a$ and voltages $\Phi_a$ are indicated in Fig.~\ref{Fig1}a.

In Eq.~\eqref{I-D-trace} the drive circuit is characterized by the
\textit{excess} part
$S^{ab}_{1}(\omega,V)=S_{ab}(\omega,V)-S_{ab}(\omega,0)$ of
current-current correlation matrix $S_{ab}(\omega,V)=\int dt\,
e^{i\omega
t}\big\langle\big\langle\delta\hat{I}_{a}(t)\delta\hat{I}_{b}(0)+
\delta\hat{I}_{b}(0)\delta\hat{I}_{a}(t)\big\rangle\big\rangle$,
which is known from the theory of quantum shot
noise~\cite{Lesovik,Buttiker,BB-review}. In particular
\begin{eqnarray}\label{S}
&&\hskip-.5cmS_{LL}(\omega,V)=\frac{2}{R_{Q}}\sum_{n}\int
d\epsilon\left[B_{RR}(\epsilon)|\t^{R}_{n}(\epsilon_{+})|^{2}
|\t^{R}_{n}(\epsilon_{-})|^{2}\right.\nonumber\\
&&\hskip-.5cm+B_{LL}(\epsilon)\big[1-\r^{*L}_{n}(\epsilon_{+})\r^{L}_{n}(\epsilon_{-})\big]
\big[1-\r^{*L}_{n}(\epsilon_{-})\r^{L}_{n}(\epsilon_{+})\big]\nonumber\\
&&\hskip-.5cm+\left.
B_{LR}(\epsilon)|\r^{L}_{n}(\epsilon_{+})|^{2}|\t^{R}_{n}(\epsilon_{-})|^{2}+
B_{RL}(\epsilon)|\t^{R}_{n}(\epsilon_{+})|^{2}|\r^{L}_{n}(\epsilon_{-})|^{2}\right]\,,
\end{eqnarray}
with the similar expressions for $S_{LR}$, $S_{RL}$ and $S_{RR}$
components, see Ref.~\cite{Buttiker} and supplementary materials for
details. Here $R_{Q}={2\pi\hbar\over e^2}$ is quantum resistance,
$|\t^{L(R)}_{n}(\epsilon)|^{2}=|\t_{n}(\epsilon+eV_{L(R)})|^{2}$ are
transmission probabilities of the drive QPC$_1$, labeled by the
transverse channel index $n$;
$|\r^{L(R)}_{n}(\epsilon)|^{2}=1-|\t^{L(R)}_{n}(\epsilon)|^{2}$ and
$V_{L}-V_{R}=V$. The statistical factors are
$B_{ab}(\epsilon)=f_{a}(\epsilon_{+})[1-f_{b}(\epsilon_{-})] +
f_{b}(\epsilon_{-})[1-f_{a}(\epsilon_{+})]$, with
$f_{L(R)}(\epsilon)=f(\epsilon+eV_{L(R)})$ being the Fermi
distributions of the two leads and $\epsilon_\pm =\epsilon\pm
\omega/2$.

The drag circuit in Eq.~\eqref{I-D-trace} is characterized by the
rectification coefficient
$\hat{\Gamma}_{2}(\omega)=\Gamma_{2}(\omega)\hat{\tau}_{z}$ of the
ac voltage fluctuations applied to the (near equilibrium) drag
QPC$_2$, where $\hat{\tau}_{z}$ is third Pauli matrix acting in the
left-right space. Rectification is given by
\begin{equation}\label{Gamma}
\Gamma_{2}(\omega)=\frac{2e}{R_{Q}}\sum_{n}\int\! d\epsilon\,
\big[f(\epsilon_{-})-f(\epsilon_{+})\big]
\Big[|\t_{n}(\epsilon_{+})|^{2}-|\t_{n}(\epsilon_{-})|^{2}\Big]\,.
\end{equation}
Characteristics of the QPC$_2$ enter through its  energy-dependent
transmission probabilities $|\t_{n}(\epsilon)|^{2}$. This expression
admits a transparent interpretation: potential fluctuations with
frequency $\omega$, say on the left of the QPC, create electron-hole
pairs with energies $\epsilon_\pm$  on the branch of right moving
particles. Consequently the electrons can pass through  the QPC with
the probability $|\t_{n}(\epsilon_{+})|^{2}$, while the holes with
the probability $|\t_{n}(\epsilon_{-})|^{2}$. The difference between
the two gives the dc current flowing across the QPC. Notice that the
energy dependence of the transmission probabilities in the drag QPC
is crucial~\cite{Footnote-2} in order to have the asymmetry between
electrons and holes, and thus non-zero rectification
$\Gamma_{2}(\omega)$.

Focusing on a single partially open channel in a smooth QPC, one may
think of the potential barrier across it as being practically
parabolic. In such a case its transmission probability is given by
\begin{equation}\label{t}
|\t(\epsilon)|^2=\left(\exp\{(eV_{g}-\epsilon)/\Delta_2\}+1\right)^{-1}\,,
\end{equation}
where $\Delta_2$ is an energy scale associated with the curvature of
the parabolic barrier in the QPC$_2$ and gate voltage $V_g$ moves
the top of the barrier relative to the Fermi energy. This form of
transmission was used  to explain QPC conductance
quantization~\cite{Glazman} and it turns out to be useful in
application to the Coulomb drag problem. Inserting \eqref{t} into
\eqref{Gamma} and carrying out energy integration, one finds
\begin{equation}\label{Gamma-1}
\Gamma_2(\omega)=\frac{2e\Delta_2}{R_{Q}}\,
\ln\left(1+\frac{\sinh^2(\omega/2\Delta_2)}
{\cosh^2(eV_{g}/2\Delta_2)}\right)
\end{equation}
for $T\ll\Delta_2$. In the other limit, $T\gg\Delta_2$, one should
replace $\Delta_2\to T$ in the Eq.~\eqref{Gamma-1}. Notice that for
small frequency $\omega\ll \Delta_2$ one has $\Gamma_2\sim
\omega^2$, making the the integral in Eq.~(\ref{I-D-trace}) to be
convergent  in $\omega\to 0$ region.

\textbf{\textit{Linear drag regime.}} For small applied voltages $V$
one expects the response current $I_D$ to be linear in $V$.
Expanding $\hat{S}_{1}(\omega,V)$ to the linear order in $V$, one
finds that only diagonal components of the current-current
correlation matrix contribute to the linear response and as a result
\begin{equation}\label{S-linear}
\hat{S}_{1}(\omega,V)=V\, \frac{\partial}{\partial\omega}
\left[\coth\frac{\omega}{2T}\right]\hat{\Gamma}_{1}(\omega)+O(V^3)\,,
\end{equation}
where $\Gamma_{1}(\omega)$ is obtained from Eq.~\eqref{Gamma} by
substituting transmission probabilities of QPC$_2$, by that of
QPC$_1$. Inserting Eq.~\eqref{S-linear} into Eq.~\eqref{I-D-trace}
one finds
\begin{equation}\label{g-D}
I_{D}=V\, \frac{R^{2}_{Q}}{4\pi}\int\!
d\omega\,\frac{\alpha_{+}(\omega)}{\omega^{2}}\,
\frac{\partial}{\partial\omega}\left[\coth\frac{\omega}{2T}\right]
\Gamma_{1}(\omega)\, \Gamma_{2}(\omega)\,,
\end{equation}
where dimensionless interaction kernel $\alpha_{+}(\omega)$ is
expressed through the components of the trans-impedance matrix as
$\alpha_{\pm}(\omega)=\big[\big(|\mathcal{Z}_{LL}|^{2}-\mathcal{Z}_{LR}\mathcal{Z}_{RL}\big)
\pm\big(|\mathcal{Z}_{RR}|^{2}-\mathcal{Z}_{LR}\mathcal{Z}_{RL}\big)\big]/2R^{2}_{Q}$.
Derived equation \eqref{g-D} has the same general structure as the
one for the drag current in bulk 2D
systems~\cite{Kamenev-Oreg,Flensberg}. Being symmetric with respect
$1\leftrightarrow 2$ permutation, it satisfies Onsager relation for
the linear response coefficient.

Assuming the load impedance of the drag circuit to be much larger
than that of the drive one $Z_{1}\ll Z_2\ll R_Q$ and the mutual
capacitance of the two circuits to be small $C_c\ll C_{R,L,s}$, see
Fig.~\ref{Fig1}a, one finds for the low frequency limit $\omega\ll
(Z_1 C_s)^{-1}$  of the interaction kernels
\begin{equation}\label{Impedance-Z0}
\alpha_{\pm}(0)=\frac{Z^2_{1}}{8R_Q^2}\, \frac{C_c^2}{C_L^2 C_R^2}
\times \left\{ \begin{array}{l}
2C_L^2 +2C_LC_R +2C_R^2 \\
C_L^2-C_R^2
\end{array} \right. \,.
\end{equation}
For $Z_{1}\to 0$ the drive QPC is shorted and the drag circuit is
insensitive to the fluctuations. Substituting now
Eq.~\eqref{Gamma-1} into Eq.~\eqref{g-D}, one finds for e.g.
low-temperature regime $T\ll \Delta_{1,2}$
\begin{equation}\label{g-D-1}
I_D= \frac{V}{R_{Q}}\,   \frac{\alpha_{+}(0)\pi^2}{6}\, {T^{2}\over
\Delta_1\Delta_2}\,  {1\over \cosh^{2}(eV_{g}/2\Delta_1)}\,,
\end{equation}
where we assumed that the gate voltage of QPC$_2$ is tuned to adjust
the top of its barrier with the Fermi energy and wrote $I_D$ as a
function of the gate voltage in QPC$_1$. We have also assumed that
$T\ll (Z_{1}C_{s})^{-1}$ to substitute $\alpha_+(\omega)$ by its dc
limit Eq.~\eqref{Impedance-Z0}. The resulting expression exhibits a
peak at $V_g=0$ similar to that depicted in Fig.~\ref{Fig1}b. Yet it
has nothing to do with the shot noise, but rather reflects
rectification of near-equilibrium thermal fluctuations (hence the
factor $T^2$) along with the electron-hole asymmetry  (hence
non-monotonous dependence on $V_g$). For monotonously increasing
functions $|\t(\epsilon)|^2$ in both circuits the linear drag is
positive (i.e. currents flow in the same direction).

\textbf{\textit{Nonlinear regime.}} At larger drive voltages drag
current ceases to be linear in $V$. Furthermore, contrary to the
linear response case, $\hat{S}_{1}(\omega,V)$ does not require
energy dependence of the transmission probabilities and could be
evaluated for energy independent $|\t_n|^2$ (this is a fare
assumption for $T,eV\ll \Delta_1$). Assuming in addition $T\ll eV$,
one finds a celebrated expression for the quantum shot
noise~\cite{Lesovik,BB-review}
\begin{equation}\label{S-nonlinear}
\hat{S}_{1}(\omega,V)=2\frac{|eV+\omega|+|eV-\omega|}{R_Q}
\sum_{n}|\t_{n}|^{2}\Big[1-|\t_{n}|^{2}\Big]\hat{\tau}_{0}\,.
\end{equation}
Inserting Eq.~\eqref{S-nonlinear} into Eq.~\eqref{I-D-trace}, after
frequency integration bounded by the voltage, one finds for the drag
current~\cite{Footnote-3}
\begin{equation}\label{I-D-nonlinear}
I_{D}=\frac{eV^2}{\Delta_{2}R_{Q}}\,  \alpha_{-}(0) \,
\sum_{n}|\t_{n}|^{2}\Big[1-|\t_{n}|^{2}\Big]\,.
\end{equation}
Here again we assumed that the detector QPC$_1$ is tuned to the
transition between the plateaus. We also assumed $eV\ll
(Z_1C_s)^{-1}$ to substitute $\alpha_-(\omega)$ by its dc value,
Eq.~(\ref{Impedance-Z0}). One should notice that while
$\alpha_{+}>0$, the sign of $\alpha_{-}$ is arbitrary. For a
completely symmetric circuit $\alpha_{-}=0$, while for extremely
asymmetric one $|\alpha_{-}|\lesssim\alpha_{+}/2$. Although we
presented derivation of Eq.~(\ref{I-D-nonlinear}) for $T\ll eV$, one
may show that it remains valid at any temperature as long as $T\ll
\mathrm{min}\{\Delta_1, (Z_1C_s)^{-1}\}$.

Equation (\ref{I-D-nonlinear}) indeed shows that the drag current is
due to the rectification of the quantum shot noise and hence
proportional to the Fano factor~\cite{Lesovik}. It again exhibits a
generic behavior depicted in Fig.~\ref{Fig1}b, but the reason is
rather different from the similar behavior in the linear regime. The
direction of the nonlinear drag current is determined by the
inversion asymmetry of the circuit (through the sign of
$\alpha_{-}$) rather than the direction of the drive current. As a
result, for a certain polarity of the drive voltage, the drag
current appears to be {\em negative}.

We discuss now a crossover between  the two regimes. Assuming that
for a generic circuit $\alpha_+\sim \alpha_-$ and comparing
Eqs.~(\ref{g-D-1}) and (\ref{I-D-nonlinear}) one concludes that the
transition from the linear to the nonlinear regime takes place at
$V\approx V^*$ with
\begin{equation}\label{V-star}
    eV^*=T^2/\Delta_1\ll T\,,
\end{equation}
for $T\ll \Delta_1$. In the opposite limit, $T>\Delta_1$, the
crossover voltage is given by the temperature $eV^*=T$. However, for
a circuit with an almost perfect inversion symmetry, i.e.
$\alpha_-\ll \alpha_+$, the nonlinear regime may be pushed to
substantially larger voltages. Such a symmetric circuitry is not
well suited for detection of the quantum shot noise.

\textbf{{\em Mesoscopic circuits.}} One or both circuits may be
represented by a multichannel quasi-1D (or 2D) mesoscopic sample. In
this case $\sum_n |\t_n(\epsilon)|^2=g(\epsilon)$ is a dimensionless
(in units of $R_Q^{-1}$) conductance of the sample as a function of
its Fermi energy. Such a conductance exhibits universal conductance
fluctuations (UCF)~\cite{UCF}, that is $g(\epsilon)=g+\delta
g(\epsilon)$, where $g\gg 1 $ is an average conductance and $\delta
g(\epsilon)\sim 1$ is a sample and energy-dependent fluctuating
part. The characteristic scale of the energy dependence of the
fluctuating part is the Thouless energy $E_{Th}=\hbar D/L^2$, where
$D$ is electronic diffusion constant and $L$ is the sample size.
Employing Eq.~(\ref{Gamma}), one finds that the rectification
 coefficient of a given mesoscopic sample may be estimated as
\begin{equation}\label{Gamma-meso}
    \Gamma(\omega)\sim \pm {e\over R_Q} \,
    \frac{\omega^2}{E_{Th}}\, ,\quad\quad \{T, \omega\}\ll E_{Th}\,.
\end{equation}
On the other hand, the nonequilibrium part of the noise correlator
Eq.~\eqref{S-nonlinear} exhibits  a well-defined average value
\begin{equation}\label{S-meso}
S_{1}(\omega,V)=2\left(|eV+\omega|+|eV-\omega|\right) {g\over
3R_Q}\,,
\end{equation}
the coefficient $1/3$ is specific to a quasi-1D geometry~\cite{BB}.

In the Coulomb drag setup, where both circuits are represented by
mesoscopic elements,    employing Eqs.~(\ref{I-D-trace}),
(\ref{g-D}) along with (\ref{Gamma-meso}), (\ref{S-meso}),  one
finds for the drag current (both linear and nonlinear)
\begin{equation}\label{g-D-meso}
I_D\sim \frac{V}{R_{Q}}\, \left( \alpha_{+} \, {T^2\over E_{Th}^2}\,
+\, \alpha_- \, {eV\over E_{Th}}\, g \right)\, ,
\end{equation}
where $T<E_{Th}$. If the load impedance of the drive circuit is
$Z_1\sim g^{-1}$, then linear in $V$ term of Eq.~(\ref{g-D-meso}) is
in agreement with the corresponding result of Ref.~\cite{Aleiner}.
The crossover between linear and nonlinear regimes takes place at
$eV^*=\alpha_+ T^2/(\alpha_- g E_{Th})$ which may be much less than
both $T$ and $E_{Th}$. As a result,  one may expect drag current to
be substantially bigger than the linear response prediction already
at the very modest bias voltage.

In summary, we have studied  Coulomb drag effect in the system of
two coupled quantum circuits. In the linear regime gate voltage
induced oscillations of the drag conductance  originate from the
particle-hole asymmetry, which is encoded in the energy dependent
transmission probabilities of the QPC. The drag conductance follows
quadratic temperature dependence at low temperatures and is peaked
at gate voltages, which correspond to the transition between QPC
conductance plateaus. Beyond the linear regime the magnitude of the
drag current is proportional to the current shot noise generated in
the drive QPC.

We are grateful to A.~Chudnovskiy, L.~Glazman, F. von-Oppen, and
B.~Shklovskii for useful discussions. We are indebted to
M.~B\"{u}ttiker for pointing out on error in a previous version of
Eq.~\eqref{S}. This work was supported by NSF grants DMR-0405212 and
DMR-0804266.


\subsection*{Supplementary materials}

The purpose of this section is to provide technical details needed
to derive Eq.~(1) of the main paper. To this end, we describe each
point contact of the quantum circuit Fig.~1a as quasi-1D adiabatic
constriction connected to two reservoirs (terminals, probes), to be
referred to as left ($L$) and right ($R$). The distribution
functions of electrons in the reservoirs of a driven circuit, are
Fermi distributions
$f_{L(R)}(\epsilon)=\big[\exp[(\epsilon+eV_{L(R)})/T]+1\big]^{-1}$,
with source-drain voltage being $V_L-V_R=V$. In the dragged circuit
distributions are assumed to be equilibrium  Fermi functions. Within
each QPC electron motion is separable into transverse and
longitudinal components. Due to the confinement transverse motion is
quantized and we assign quantum number $n$ to label transverse
conduction channels with $\phi_{n}(r_{\perp})$ being corresponding
transversal wave function. The longitudinal motion is describe in
terms of the extended scattering states -- normalized electron plane
waves incident from the left
\begin{equation}\label{u-L}
u^{L}_{n}(k,r)=\frac{\phi_{n}(r_{\perp})}{\sqrt{v}}\left\{
\begin{array}{cc}
e^{ikx}+\r_{n}(k)e^{-ikx} & x\to-\infty \\
\t_{n}(k)e^{ikx} & x\to+\infty
\end{array}
\right.
\end{equation}
and right
\begin{equation}\label{u-R}
u^{R}_{n}(k,r)=\frac{\phi_{n}(r_{\perp})}{\sqrt{v}}\left\{
\begin{array}{cc}
e^{-ikx}+\r_{n}(k)e^{ikx} & x\to+\infty \\
\t_{n}(k)e^{-ikx} & x\to-\infty
\end{array}
\right.
\end{equation}
onto mesoscopic scattering region. Here $k$ and $v$ are the electron
wave vector and velocity, $\t_{n}(k)$ and $\r_{n}(k)$ are channel
specific transmission and reflection amplitudes. Second quantized
electron field operator is introduced in the standard way
\begin{equation}
\hat{\Psi}(r,t)=\sum_{nk}\left[\hat{\psi}^{L}_{n}(k,t)u^{L}_{n}(k,r)+
\hat{\psi}^{R}_{n}(k,t)u^{R}_{n}(k,r)\right]\,,
\end{equation}
where $\hat{\psi}^{L(R)}_{n}(k,t)$ are fermion destruction operators
in the left and right reservoirs correspondingly. For the future use
we define also current operator
\begin{equation}\label{Current}
\hat{I}_{a}(t)=\sum_{nk}M^{a}_{nn'}\hat{\psi}^{\dag
a}_{n}(k,t)\hat{\psi}^{a}_{n'}(k',t)\,,
\end{equation}
which has matrix elements $M^{a}_{nn'}=(e/2im)\int
dr_{\perp}\big[u^{*a}_{n}(k,r)\partial_{x}u^{a}_{n'}(k',r)-
\partial_{x}[u^{*a}_{n}(k,r)]u^{a}_{n'}(k',r)\big]$,
constructed from the scattering states \eqref{u-L}--\eqref{u-R}.
Based on the orthogonality condition of transverse wave functions
$\int
dr_{\perp}\phi^{\phantom{*}}_{n}(r_{\perp})\phi^{*}_{n'}(r_{\perp})=\delta_{nn'}$,
direct calculation gives
\begin{equation}\label{M}
M^{L}_{nn'}=-e\delta_{nn'}\left(\begin{array}{ll}
\r^{*}_{n}(k)\r_{n}(k')-1 & \r^{*}_{n}(k)\t_{n}(k') \\
\t^{*}_{n}(k)\r_{n}(k') & \t^{*}_{n}(k)\t_{n}(k')
\end{array}\right)\,,
\end{equation}
and a similar result for $M^{R}_{nn'}$. In Eq.~\eqref{M} we have
suppressed phase factors $e^{\pm i(k-k')x}\approx 1$, since
$|k-k'|\sim L_T^{-1}\ll x^{-1}$,  and the coordinate $x$ is confined
by the sample size $L\ll L_T$. On the other hand, we do not keep
fast oscillating factors $e^{\pm 2ik_Fx}$, since $x\sim L\gg
k_F^{-1}$. However, one must keep explicitly momentum (or
equivalently energy) dependence for transmission amplitudes, which
translates later into particle--hole asymmetry factor
$\Gamma(\omega)$.

Dynamics of $\hat{\psi}^{L(R)}_{n}(k,t)$ operators is governed by
the action
\begin{equation}\label{Action-0}
i\mathcal{S}_{0}=\int_{C}dt\sum_{ja}\sum_{nk}\bar{\psi}^{a}_{jn}(k,t)
i\hat{\mathbf{G}}^{-1}\psi^{a}_{jn}(k,t)
\end{equation}
defined along the Keldysh contour, where
$\hat{\mathbf{G}}^{-1}=(i\partial_{t}-\xi_k+eV_{a})$ is Green's
function operator with $\xi_k$ being electron energy. Additional
subscript $j$ in Eq.~\eqref{Action-0} labels drive $(j=1)$ and drag
$(j=2)$ QPC's. As usual for Keldysh technique one splits time
integration into forward and backward pathes and replicates each
fermion field into two components
$\psi\Rightarrow\psi^{\rightleftarrows}$, which belong now to the
different contour branches. It is convenient also to perform Keldysh
rotation
\begin{equation}
\left(\begin{array}{c} \psi^{+} \\
\psi^{-}\end{array}\right)=\hat{\mathbb{L}}\hat{\sigma}_{z}\left(\begin{array}{c} \psi^{\rightarrow} \\
\psi^{\leftarrow}\end{array}\right),\,
\left(\begin{array}{c} \bar{\psi}^{+} \\
\bar{\psi}^{-}\end{array}\right)=\hat{\mathbb{L}}\left(\begin{array}{c} \bar{\psi}^{\rightarrow} \\
\bar{\psi}^{\leftarrow}\end{array}\right)\,,\hat{\mathbb{L}}=\frac{1}{\sqrt{2}}
\left(\begin{array}{cc}1&-1\\1&\phantom{-}1\end{array}\right),
\end{equation}
where $\hat{\sigma}_z$ is third Pauli matrix acting in the Keldysh
space. In this rotated basis quadratic action \eqref{Action-0} gives
following electron correlators
\begin{equation}\label{G}
-i\langle\langle\psi^{a\alpha}_{in}(k,t)\bar{\psi}^{b\beta}_{jn}(k,t')\rangle\rangle
=\delta_{ij}\delta_{ab}\mathbf{G}^{\alpha\beta}_{ia}(t,t')\,,
\end{equation}
where Keldysh Green's function matrix has familiar triangular
structure
\begin{equation}\label{G-matrix}
\mathbf{G}^{\alpha\beta}_{ia}(t,t')=\left(
\begin{array}{cc}
G^{R}_{ia}(t,t') & G^{K}_{ia}(t,t') \\ 0 & G^{A}_{ia}(t,t')
\end{array}
\right)^{\alpha\beta}.
\end{equation}
Retarded/Advanced/Keldysh components of $\mathbf{G}^{\alpha\beta}$
are given by
\begin{eqnarray}\label{G-RAK}
&&iG^{R}_{ia}(t,t')=\theta(t-t')e^{-i(\xi_{k}+eV_{a})(t-t')}\,,\nonumber\\
&&iG^{A}_{ia}(t,t')=-\theta(t'-t)e^{-i(\xi_{k}+eV_{a})(t-t')}\,,\\
&&iG^{K}_{ia}(t,t')=[1-2f_{a}(\epsilon_k)]e^{-i(\xi_{k}+eV_{a})(t-t')}\,.\nonumber
\end{eqnarray}

Having described quantum point contacts individually we introduce
now the interaction term between them
\begin{equation}\label{Action-int}
i\mathcal{S}_{\mathrm{int}}=\sum_{ab\alpha\beta}\iint^{+\infty}_{-\infty}dtdt'\,
I^{\alpha}_{1a}(t)\mathbf{K}^{\alpha\beta}_{ab}(t-t')I^{\beta}_{2b}(t')\,.
\end{equation}
Here $I_{jR(L)}(t)$ are current operators \eqref{Current}, on the
right (left) of the QPC$_j$, coupled by the kernel
$\mathbf{K}(t-t')$, which encodes electromagnetic environment of the
circuit. Interaction kernel retarded and advanced components are
directly related to the trans-impedance matrix of the circuit
$\mathbf{K}^{R(A)}_{ab}(\omega)=\mathcal{Z}^{R(A)}_{ab}(\omega)/(\omega\pm
i0)$, while Keldysh component can be restored from the
fluctuation-dissipation theorem:
$\mathbf{K}^{K}_{ab}(\omega)=[\mathbf{K}^{R}_{ab}(\omega)-
\mathbf{K}^{A}_{ab}(\omega)]\coth(\omega/2T)$, i.e. we assume the
surrounding electric environment to be close to equilibrium.

\begin{figure}
  \includegraphics[width=8cm,height=7cm]{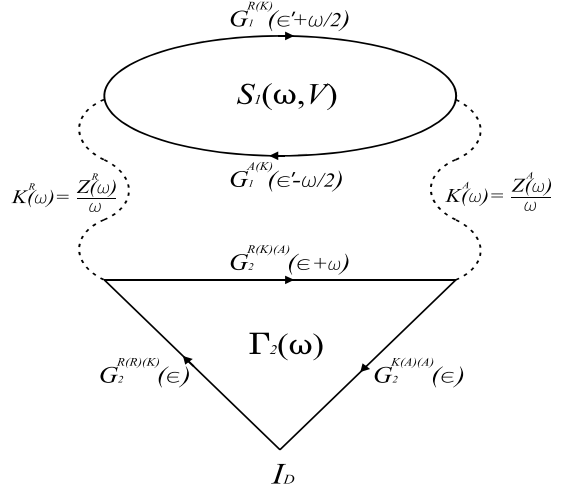}
  \caption{Drag current $I_D$ in the second order in
  inter-circuit interactions $\mathbf{K}=\mathcal{Z}/\omega$ (wavy lines). The drag circuit
  is represented by triangular rectification vertex $\Gamma_2(\omega)$, while the drive circuit by the
  non-equilibrium current-current correlator $S_1(\omega, V)$ (loop).  }\label{Fig2}
\end{figure}

Within this formalism, drag current is found by averaging $I_{2}$
over the  fermionic degrees of freedom
\begin{equation}\label{I-D-def}
I_{D}=
\int\mathbf{D}[\psi\bar{\psi}]\,\mathrm{Tr}\left[\bar{\psi}_{2}M\psi_{2}\right]\,
\exp\big(i\mathcal{S}[\bar{\psi}\psi]\big)\,,
\end{equation}
where $\mathcal{S}=\mathcal{S}_{0}+\mathcal{S}_{\mathrm{int}}$ and
we used  expression  \eqref{Current} for the current operator. Being
interested in the leading perturbative result, we expand
$\exp(i\mathcal{S})$ to the second order in interaction term
$\mathcal{S}_{\mathrm{int}}$. This way one obtains
\begin{eqnarray}\label{I-D-def-perturb}
\!\!\!\!\!I_{D}=\frac{1}{2}\int\mathbf{D}[\psi\bar{\psi}]\!\!\!\!\!
&&\mathrm{Tr}\left[\bar{\psi}_{2}M\psi_{2}\right]\nonumber\\
&&\mathrm{Tr}\left[I_{1}\mathbf{K}I_{2}\right]
\mathrm{Tr}\left[I_{1}\mathbf{K}I_{2}\right]
\exp\big(i\mathcal{S}_{0}[\bar{\psi}\psi]\big).
\end{eqnarray}
Remaining Gaussian integral may be evaluating using the Wick's
theorem. One inserts expressions \eqref{Current} for the current
operators into the traces of \eqref{I-D-def-perturb} and takes into
the account all the possible Wick's contraction between
$\psi$-fields. The latter are given by the Green's functions
\eqref{G}. This way we find our main result for the drag current
[Eq.~\eqref{I-D-trace}] shown diagrammatically in the
Fig.~\ref{Fig2}. The interaction kernels $\alpha_{\pm}(\omega)$,
rectification coefficient $\Gamma_{2}(\omega)$ and noise power
$S_{ab}(\omega,V)$ are given explicitly by the following Keldysh
traces:
\begin{eqnarray}
&&\alpha_{+(-)}(\omega)=\frac{1}{2R^{2}_{Q}}
\mathrm{Tr}\left[\hat{\mathcal{Z}}^{R}(\omega)\hat{\tau}_{0(z)}
\hat{\mathcal{Z}}^{A}(\omega)\tau_{z}\right]\,,\label{traces-alpha}\\
&&S_{ab}(\omega,V)=\mathrm{Tr}\left[\hat{\mathbf{G}}\hat{\sigma}_{x}\hat{M}_{a}
\hat{\mathbf{G}}\hat{\sigma}_{x}\hat{M}_{b}\right]\,,\label{traces-S}\\
&&\Gamma_{2}(\omega)=\mathrm{Tr}\left[\hat{\mathbf{G}}\hat{\sigma}_{x}\hat{M}
\hat{\mathbf{G}}\hat{\sigma}_{0}\hat{M}\hat{\mathbf{G}}\hat{\sigma}_{0}
\hat{M}\right]\,,\label{traces-Gamma}
\end{eqnarray}
where $\hat{\tau}$ and $\hat{\sigma}$ are two sets of  Pauli
matrices in left-right and Keldysh spaces, correspondingly.

Remaining steps of algebra concern calculation of the traces in
Eqs.~\eqref{traces-alpha}--\eqref{traces-Gamma}. For the interaction
coefficients $\alpha_{\pm}$ it is straightforward. Let us
demonstrate how Eq.~(2) for $S_{LL}$ is recovered from
Eq.~\eqref{traces-S}. To this end we insert current matrix $M^{L}$
[Eq.~\eqref{M}] along with
$\hat{\mathbf{G}}=\mathrm{diag}\{\hat{G}_{L},\hat{G}_{R}\}$ into
Eq.~\eqref{traces-S} and calculate trace over left-right subspace,
which gives
\begin{eqnarray}\label{traces-S-1}
&&\hskip-.2cm
S_{LL}(\omega,V)=e^2\sum_{n}\int\frac{d\epsilon}{2\pi}\int\frac{d\xi
d\xi'}{4\pi^2}\\
&&\hskip-.2cm\left[\Tr\big[\hat{G}_{L}(\epsilon_{+},\xi)
\hat{\sigma}_{x}\hat{G}_{L}(\epsilon_{-},\xi')\hat{\sigma}_{x}\big]
[\r^{*}_{n}(\xi)\r_{n}(\xi')-1][\r^{*}_{n}(\xi')\r_{n}(\xi)-1]\right.\nonumber\\
&&\hskip-.2cm+\Tr\big[\hat{G}_{L}(\epsilon_{+},\xi)
\hat{\sigma}_{x}\hat{G}_{R}(\epsilon_{-},\xi')\hat{\sigma}_{x}\big]
|\r_{n}(\xi)|^{2}|\t_{n}(\xi')|^{2}\nonumber\\
&&\hskip-.2cm+\Tr\big[\hat{G}_{R}(\epsilon_{+},\xi)
\hat{\sigma}_{x}\hat{G}_{L}(\epsilon_{-},\xi')\hat{\sigma}_{x}\big]
|\t_{n}(\xi)|^{2}|\r_{n}(\xi')|^{2}\nonumber\\
&&\hskip-.2cm\left.+\Tr\big[\hat{G}_{R}(\epsilon_{+},\xi)
\hat{\sigma}_{x}\hat{G}_{R}(\epsilon_{-},\xi')\hat{\sigma}_{x}\big]
|\t_{n}(\xi)|^{2}|\t_{n}(\xi')|^{2}\nonumber\right]\,.
\end{eqnarray}
Recall here that $\hat{G}_{L(R)}$ are still matrices in the Keldysh
subspace. Using Eq.~\eqref{G-matrix} one calculates remaining traces
over the Keldysh subspace
\begin{eqnarray}\label{traces-S-2}
\mathrm{Tr}\left[\hat{G}_{a}\hat{\sigma}_{x}\hat{G}_{b}\hat{\sigma}_{x}\right]=
G^{R}_{a}G^{A}_{b}+G^{A}_{a} G^{R}_{b} +G^{K}_{a}G^{K}_{b}\,,
\end{eqnarray}
and performs final integration with the help of Green's functions
$G^{R(A)}_{a}(\epsilon,\xi)=(\epsilon-\xi+eV_{a}\pm i0)^{-1}$ and
$G^{K}_{a}(\epsilon,\xi)=-2\pi
i\delta(\epsilon-\xi+eV_{a})\big[1-2f_{a}(\xi)\big]$, which follows
from Fourier transforms of Eq.~\eqref{G-RAK}. It is not difficult to
see now that each Keldysh trace in Eq.~\eqref{traces-S-1} defines
statistical occupation factors $B_{ab}(\epsilon)$ used in Eq.~(2),
namely
$\Tr\left[\hat{G}_{a}\hat{\sigma}_{x}\hat{G}_{b}\hat{\sigma}_{x}\right]\propto
B_{ab}(\epsilon)$. As a result, collecting all the factors, one
finds from Eq.~\eqref{traces-S-2} the final form of noise power
given by Eq.~(2). In complete analogy one may calculate $S_{RR}$ and
$S_{LR}$ components of the noise power:
\begin{eqnarray}
&&\hskip-.5cm S_{RR}(\omega,V)=\frac{2}{R_{Q}}\sum_{n}\int
d\epsilon\left[B_{LL}(\epsilon)|\t^{L}_{n}(\epsilon_{+})|^{2}
|\t^{L}_{n}(\epsilon_{-})|^{2}\right. \\
&&\hskip-.5cm+B_{RR}(\epsilon)\big[1-\r^{*R}_{n}(\epsilon_{+})\r^{R}_{n}(\epsilon_{-})\big]
\big[\r^{*R}_{n}(\epsilon_{-})\r^{R}_{n}(\epsilon_{+})-1\big]\nonumber\\
&&\hskip-.5cm\left.+B_{LR}(\epsilon)|\t^{L}_{n}(\epsilon_{+})|^{2}|\r^{R}_{n}(\epsilon_{-})|^{2}
+B_{RL}(\epsilon)|\r^{R}_{n}(\epsilon_{+})|^{2}|\t^{L}_{n}(\epsilon_{-})|^{2}\right]\nonumber\,,
\end{eqnarray}
\begin{eqnarray}
&&\hskip-.5cm S_{LR}(\omega,V)=-\frac{2}{R_{Q}}\sum_{n}\int
d\epsilon
\\
&&\hskip-.5cm
\left[B_{LL}(\epsilon)\big[\t^{*L}_{n}(\epsilon_{+})\t^{L}_{n}(\epsilon_{-})-
\t^{*L}_{n}(\epsilon_{+})\t^{L}_{n}(\epsilon_{-})\r^{*L}_{n}(\epsilon_{-})
\r^{L}_{n}(\epsilon_{+})\big]\right.\nonumber\\
&&\hskip-.5cm
+B_{RR}(\epsilon)\big[\t^{R}_{n}(\epsilon_{+})\t^{*R}_{n}(\epsilon_{-})-
\t^{*R}_{n}(\epsilon_{+})\t^{R}_{n}(\epsilon_{-})\r^{R}_{n}(\epsilon_{+})
\r^{*R}_{n}(\epsilon_{-})\big]\nonumber \\
&&\hskip-.5cm
+B_{LR}(\epsilon)\t^{*L}_{n}(\epsilon_{+})\t^{R}_{n}(\epsilon_{-})
\r^{*R}_{n}(\epsilon_{-})\r^{L}_{n}(\epsilon_{+})\nonumber \\
&&\hskip-.5cm \left. +B_{RL}(\epsilon)
\t^{*R}_{n}(\epsilon_{+})\t^{L}_{n}(\epsilon_{-})
\r^{*L}_{n}(\epsilon_{-})\r^{R}_{n}(\epsilon_{+})\right]\nonumber\,.
\end{eqnarray}
Notice that cross-correlation component is negative.

Finding $\Gamma_2(\omega)$ one uses the fact that Green's function
is diagonal unity in the left-right subspace
$\hat{\mathbf{G}}=\hat{G}\hat{\tau}_{0}$ and faces Keldysh trace of
the kind
\begin{eqnarray}
\mathrm{Tr}\left[\hat{G}\hat{\sigma}_{x}\hat{G}\hat{\sigma}_{0}
\hat{G}\hat{\sigma}_{0}\right]
=\sum_{\pm}\left[G^{R}(\epsilon)G^{R}(\epsilon\pm\omega)G^{K}(\epsilon)\right.\nonumber\\
+\left.G^{R}(\epsilon)G^{K}(\epsilon\pm\omega)G^{A}(\epsilon)+
G^{K}(\epsilon)G^{A}(\epsilon\pm\omega)G^{A}(\epsilon)\right].
\end{eqnarray}
To simplify this equation one should decompose each Keldysh
component of the Green's function using fluctuation-dissipation
relation
$G^{K}(\epsilon)=\big[G^{R}(\epsilon)-G^{A}(\epsilon)\big][1-2f(\epsilon)]$
and keep in the resulting expression only those terms, which have
different causality. Combinations having three Green's functions of
the same kind, like $G^{A}G^{A}G^{A}$ and $G^{R}G^{R}G^{R}$, will
not contribute. This way, one finds that
$\mathrm{Tr}\left[\hat{G}\hat{\sigma}_{x}\hat{G}
\hat{\sigma}_{0}\hat{G}\hat{\sigma}_{0}\right]
\propto\big[f(\epsilon_{-})-f(\epsilon_{+})\big]$, with
$\epsilon_{\pm}=\epsilon\pm\omega/2$. Remaining trace over the
current vertex matrices $M_2$ reduces to the transmission
probabilities at shifted energies, namely
$\mathrm{Tr}\big[M_{2}M_{2}M_{2}\big]\propto\pm|\t_{n}(\epsilon_{\pm})|^{2}$.
As the result, imposing remaining $\epsilon$ integration and
summation over transverse channels, one arrives at
$\Gamma_{2}(\omega)$ in the form of Eq.~(4) of the main text. This
completes our derivation.


\begin{thebibliography}{99}
\bibitem{Footnote-1} This is the case in the lowest (second) order in inter-circuit interactions.
In higher orders in interactions drag conductance may be temperature
independent, see  Ref.~\cite{LK}.
\bibitem{Footnote-2} We neglect here a small curvature of the particles dispersion relation.
The later is the sole reason for the drag in bulk 2D systems. In
quantum circuits its effect is small as $\Delta/\epsilon_F\ll 1$.
\bibitem{Footnote-3} Complete form of Eq.~\eqref{I-D-nonlinear} contains
also an additional term
$$
\hskip 0.6cm\delta
I_{D}(V)=\int\frac{d\omega}{4\pi}\frac{\mathcal{Z}_{LL}-\mathcal{Z}_{RR}}{\omega^{2}}
\left[\mathcal{Z}_{RL}\delta S^{LR}_{1}+\mathcal{Z}_{LR}\delta
S^{RL}_{1}\right]\Gamma_{2}(\omega)\,.
$$
It does not contribute neither to the linear nor to the nonlinear
response regimes discussed in the text. Due to the symmetry of
$\delta S^{LR}_{1}$ and $\delta S^{RL}_{1}$ with respect to the
change $V\to-V$, at small voltages $\delta I_{D}\propto V^{2}$ in
contrast to $I_D\propto V$ [Eq.~\eqref{g-D-1}]. For the nonlinear
regime, when energy dependence of the transmissions can be negleced,
$\delta S^{LR}_{1}=\delta S^{RL}_{1}$, and $\delta I_{D}$ vanishes,
since $\mathcal{Z}_{LR}=-\mathcal{Z}_{RL}$.
\end{thebibliography}
\end{document}